\documentclass{pnastwo}

\usepackage[dvips]{graphicx}
\usepackage{pnastwoF}
\usepackage{amssymb,amsfonts,amsmath}

\url{www.pnas.org/cgi/doi/10.1073/pnas.0709640104}
\copyrightyear{2008}
\issuedate{Issue Date}
\volume{Volume}
\issuenumber{Issue Number}

\begin{document}

\title{Missing and spurious interactions\\ and the reconstruction of complex networks}

\author{Roger Guimer\`a\affil{1}{Department of Chemical and Biological Engineering, Northwestern University, Evanston, IL, US}\affil{2}{Northwestern Institute on Complex Systems (NICO), Northwestern University, Evanston, IL, US}
\and
Marta Sales-Pardo\affil{1}{}\affil{2}{}\affil{3}{Northwestern University Clinical and Translational Science Institute (NUCATS), Northwestern University, Chicago, IL, US}}

\contributor{Submitted to Proceedings of the National Academy of Sciences
of the United States of America}

\maketitle

\begin{article}
\begin{abstract}
	Network analysis is currently used in a myriad of contexts: from identifying potential drug targets to predicting the spread of epidemics and designing vaccination strategies, and from finding friends to uncovering criminal activity. Despite the promise of the network approach, the reliability of network data is a source of great concern in all fields where complex networks are studied. Here, we present a general mathematical and computational framework to deal with the problem of data reliability in complex networks. In particular, we are able to reliably identify both missing and spurious interactions in noisy network observations. Remarkably, our approach also enables us to obtain, from those noisy observations, network reconstructions that yield estimates of the true network properties that are more accurate than those provided by the observations themselves. Our approach has the potential to guide experiments, to better characterize network data sets, and to drive new discoveries.
\end{abstract}


\abbreviations{BM, stochastic block model; HRG, hierarchical random graph}



\dropcap{T}he structure of the network of interactions between the units of a system affects the system's dynamics, and conveys information about the functional needs of the system, its evolution, and the role of individual units. For these reasons, network analysis has become a cornerstone of fields as diverse as systems biology and sociology~\cite{amaral04}. Unfortunately, the reliability of network data is often a source of concern. In systems biology, high-throughput technologies hold the promise to uncover the intricate processes within the cell, but are also reportedly inaccurate. Protein interaction data provide, arguably, the most blatant example of data inaccuracy: in 2002, a systematic comparison of several high-throughput methods to a reference high-quality data set showed that these methods have accuracies below 20\%~\cite{mering02}. Additionally, different methods result in networks that have different topological properties~\cite{yu08}, and the coverage of real interactomes is very limited: 80\% of the interactome of yeast~\cite{yu08} and 99.7\% of the human interactome~\cite{stumpf08,amaral08} are still unknown.

In the social sciences, missing data due to individual non-response and dropout~\cite{schafer02}, informant inaccuracy~\cite{butts03}, and sampling biases~\cite{kossinets06a} are also pervasive. Simulation studies have established that these inaccuracies can lead to fundamentally wrong estimates of network properties and to misleading conclusions~\cite{kossinets06a}, which is particularly worrisome at a time when social network analysis is being used for finding new friends and partners, singling out key individuals in organizations, and identifying criminal activity.


Despite these concerns, the issue of network reliability has only been addressed in a field-by-field basis (for example, to deal with protein-protein interactions~\cite{jansen03,chiang09} or to take into account informant inaccuracy in social networks~\cite{butts03}), and in studies that only address parts of the problem (for example, to detect missing interactions~\cite{clauset08}). Therefore, a general framework to deal with the problem of data reliability in complex networks is lacking. Here, we develop such a framework. Specifically, we show that within our framework we can reliably: (i) identify false negatives (missing interactions) and false positives (spurious interactions), and (ii) generate, from a single observed network, a reconstructed network whose properties (clustering coefficient, modularity, assortativity, epidemic spreading threshold, and synchronizability, among others) are closer to the ``true'' underlying network than those of the observed network itself. We show that our approach outperforms previous attempts to predict missing and spurious interactions, and illustrate the potential of our method by applying it to a protein interaction network of yeast~\cite{gavin06}. We end by discussing how our approach will help to guide experiments and new discoveries, and to better characterize important data sets.

\section{General reliability formalism}
Consider an observed network with adjacency matrix $A^O$; $A^O_{ij}=1$ if nodes $i$ and $j$ are connected and $0$ otherwise. We assume that this observed network is a realization of an underlying probabilistic model, either because the network itself is the result of a stochastic process, because the measurement has uncertainty, or both~\cite{butts03}~\footnote{For simplicity, in this manuscript we use language that is consistent with a situation in which a true network exists but is obscured by the inaccuracies of the observation process. Thus, we talk about the ``true'' network, which has no ``errors,'' and about ``observed'' networks, which have ``errors.'' However, the formalism is valid even if the network is itself the outcome of a stochastic process.}. Let us call $\mathcal{M}$ the set of generative models that could conceivably give rise to the observed network, and $p(M|A^O)$ the probability that $M\in\mathcal{M}$ is the model that gave rise to the observation $A^O$. If we could get a new observation of the network, the outcome would in general be different from $A^O$; our best estimate for the probability $p(X=x)$ for an arbitrary network property $X$ is
\begin{equation}
	p(X=x|A^O) = \int_{\mathcal{M}} dM \, p(X=x|M) \, p(M|A^O)\,,
	\label{e-basic}
\end{equation}
where $p(X=x|M)$ is the probability that $X=x$ in a network generated with model $M$. Using Bayes theorem, we can rewrite Eq.~(\ref{e-basic}) as
\begin{equation}
	 p(X=x|A^O) = \frac{\int_{\mathcal{M}} dM \, p(X=x|M) \, p(A^O|M) \, p(M)}{\int_{\mathcal{M}} dM' \, p(A^O|M')p(M')}\,,
	\label{e-bayes}
\end{equation}
where $p(A^O|M)$ is the probability that model $M$ gives rise to $A^O$ among all possible adjacency matrices, and $p(M)$ is the a priori probability that model $M$ is the correct one. We call $p(X=x|A^O)$ the reliability of the $X=x$ measurement.

\section{Stochastic block models}
Given the generality of these arguments, the key to good estimates of reliability is to identify sets of models that are general, empirically grounded, and analytically or computationally tractable. Here, we focus on the family $\mathcal{M}_{\rm BM}$ of stochastic block models~\cite{white76,holland83}. In a stochastic block model, nodes are partitioned into groups and the probability that two nodes are connected depends only on the groups to which they belong (Fig.~\ref{f-blockmodel}).

Stochastic block models are empirically grounded in that they capture two ubiquitous and fundamental properties of real complex networks. First, nodes in real networks are often organized into modules or communities~\cite{girvan02,guimera07}, which may overlap~\cite{sawardecker09} or be hierarchically nested~\cite{sales-pardo07,arenas06,clauset08}, so that connections are relatively more abundant within modules than between modules. In most real-world networks, this modularity is significantly larger than expected from chance~\cite{guimera04,guimera07}. Second, nodes in real networks fulfill distinct roles and connect to each other depending on these roles~\cite{white76,guimera07}. Role-to-role connections are not necessarily assortative \cite{newman03h,guimera07}, that is, nodes with a certain role may or may not tend to connect with other nodes with the same role. In general, stochastic block models are particularly appropriate when nodes belong to groups and interact with each other depending on their group membership (regardless of whether interactions occur mostly within groups or between groups).

Stochastic block models are also appropriate in that they can capture other more general connectivity correlations in the network. For example, if people establish social connections with others according to age, then a block model that partitions individuals into age groups will capture some of the correlations in the network.

In general, complex networks result from a combination of mechanisms, including modularity, role structure, and maybe other factors. Although partitions into modules, roles, and age groups, for example, can be very different from each other, some block model in the $\mathcal{M}_{\rm BM}$ family is likely to capture each of them separately; by sampling over all models $M \in \mathcal{M}_{\rm BM}$ we capture a variety of correlations, ideally to the exact degree that they are relevant.

Additionally, stochastic block models are analytically tractable because in a stochastic block model the probability that nodes $i$ and $j$ are connected depends only on the groups to which they belong~\cite{holland83}. Therefore, we can calculate the reliability of individual links and the reliability of entire networks.

\section{Link reliability: missing and spurious interactions}
The reliability of an individual link is $R^L_{ij} \equiv p_{\rm BM}(A_{ij}=1|A^O)$, that is, the probability that the link ``truly'' exists given our observation of the whole network (and our choice of the family of stochastic block models). Assuming no prior knowledge about the suitability of the models, we obtain (Methods)
\begin{equation}
	R^L_{ij} = \frac{1}{Z} \sum_{P \in \mathcal{P}} \left( \frac{l^O_{\sigma_i \sigma_j} + 1}{r_{\sigma_i \sigma_j} + 2} \right) \exp [-\mathcal{H}(P)] \, ,
	\label{e-plink}
\end{equation}
where the sum is over partitions $P$ in the space $\mathcal{P}$ of all possible partitions of the network into groups, $\sigma_i$ is node $i$'s group (in partition $P$), $l^O_{\alpha\beta}$ is the number of links in the observed network between groups $\alpha$ and $\beta$, and $r_{\alpha\beta}$ is the maximum possible number of links between $\alpha$ and $\beta$ (Fig.~\ref{f-blockmodel}). The function $\mathcal{H}(P)$ is a function of the partition
\begin{equation}
	\mathcal{H}(P) = \sum_{\alpha \le \beta} \left[ \ln (r_{\alpha \beta} + 1) + \ln \binom{r_{\alpha \beta}}{l^O_{\alpha \beta}} \right] \,,
	\label{e-H}
\end{equation}
and $Z=\sum_{P \in \mathcal{P}} \exp [-\mathcal{H}(P)]$.

In practice, it is not possible to sum over all partitions even for small networks \footnote{The number of distinct partitions of $N$ elements into groups is $\sum_{k=1}^{N}\frac{1}{k!} \sum_{l=1}^k \binom{k}{l} \, (-1)^{k-l} \, l^N$, which grows faster than any finite power of $N$.}. However, since Eq. (\ref{e-plink}) has the same mathematical form as an ensemble average in statistical mechanics~\cite{huang87}, one can use the Metropolis algorithm to correctly sample relevant partitions (that is, partitions that significantly contribute to the sum) and obtain estimates for the link reliability (Methods).

We use the link reliability to identify missing and spurious interactions in network observations. We evaluate the performance of our approach using five high-quality networks: the social network of interactions between people in a karate club~\cite{zachary77}, the social network of frequent associations between 62 dolphins~\cite{lusseau03}, the air transportation network of Eastern Europe~\cite{guimera05b}, the neural network of the nematode {\it C. elegans}~\cite{white86}, and the metabolic network of {\it E. coli}~\cite{reed03,guimera07a}. All of these networks have been manually curated and are widely used in the literature as model systems. Therefore, in what follows we assume that each of these networks is the ``true'' network $A^T$ and is error-free. We then generate hypothetical observations $A^O$ by adding or removing random connections from $A^T$, and evaluate the ability of our approach to recover the features of the true network~\footnote{By adding and removing connections in this way, we are implicitly focusing on random errors; we discuss at the end how our approach can also deal with systematic (or, in general, correlated) errors.}.

To quantitatively study missing interactions, we generate observed networks $A^O$ by removing random links from the true network $A^T$. We then estimate the link reliability $R_{ij}^L$ for each of these false negatives ($A^O_{ij}=0$ and $A^T_{ij}=1$), as well as for the true negatives ($A^O_{ij}=0$ and $A^T_{ij}=0$). We measure the algorithm's ability to identify missing interactions by ranking the reliabilities (in decreasing order) and calculating the probability that a false negative has a higher ranking than a true negative~\cite{clauset08}. Similarly, we quantify the ability to identify spurious interactions by adding random links to the true network, obtaining and ranking the link reliabilities (again, in decreasing order), and calculating the probability that a false positive ($A^O_{ij}=1$ and $A^T_{ij}=0$) is ranked lower than a true positive ($A^O_{ij}=1$ and $A^T_{ij}=1$).

In Fig.~\ref{f-mis_bog_performance}, we compare our approach to the hierarchical random graph (HRG) approach of Clauset {\it et al.}~\cite{clauset08} and to a local algorithm based on the number of common neighbors between each pair of nodes~\cite{liben-nowell07,clauset08} (Methods; see Supporting Information, Fig. S1, for a comparison to other local algorithms). We find that, except for one network, our approach consistently outperforms all others at identifying both missing interactions and spurious interactions. Our approach is also the only one that performs consistently well for all networks (unlike local algorithms, which work well for some networks but very poorly for others) and for both missing and spurious interactions (unlike the HRG algorithm, which performs comparatively worse at detecting spurious interactions\footnote{A plausible explanation for this behavior is that, because in the HRG model most parameters are used to ``fit'' low-level features of the network (pairs of nodes, triplets of nodes, and so on), the HRG approach may overfit spurious links.}). Our algorithm is also consistently the most accurate when applied to a number of model networks, including networks with hierarchically nested modules, networks with a strongly disassortative role structure, and non-modular scale-free networks (Supporting Information, Fig. S2). We find that only when the network is strictly a hierarchical random graph, is the HRG approach slightly more accurate at predicting missing interactions than the BM approach (Supporting Information Secs. 2 and 3). Remarkably, even for strict hierarchical random graphs, the BM approach is more accurate at identifying spurious interactions.

\section{Network reliability and network reconstruction}

The success at detecting both missing and spurious interactions confirms that our approach is able to uncover the structural features of the true network $A^T$. The natural question is thus whether it is possible to ``reconstruct'' the observation $A^O$ to gain greater insight into the global structure of $A^T$. This is difficult because, in general, adding a few candidate missing interactions and removing a few candidate spurious interactions does not give satisfactory network reconstructions (one of the main problems being that one does not know, a priori, how many missing and spurious interactions there are).

Therefore, the first step toward network reconstruction is to obtain the network reliability $R^N_A \equiv p_{\rm BM}(A|A^O)$, that is, the probability that $A$ is the true network given our observation $A^O$ (and our choice of the family of stochastic block models). We obtain (Methods)
\begin{equation}
	R^N_A = \frac{1}{Z} \sum_{P \in \mathcal{P}} h(A; A^O, P) \exp [-\mathcal{H}(P)] \, ,
	\label{e-pnet}
\end{equation}
where
\begin{eqnarray}
	h(A; A^O, P) & = & \exp \left\lbrace \sum_{\alpha \le \beta} \left[ \ln \left( \frac{r_{\alpha \beta} + 1}{2 \, r_{\alpha \beta} + 1} \right)\right.\right.\nonumber\\
	& + & \left.\left. \ln \left( \frac{\binom{r_{\alpha \beta}}{l^O_{\alpha \beta}}}{\binom{2 \, r_{\alpha \beta}}{l_{\alpha \beta} + l^O_{\alpha \beta}}} \right) \right] \right\rbrace \;,
	\label{e-pnet_h}
\end{eqnarray}
$l_{\alpha \beta}$ is the number of links in $A$ between groups $\alpha$ and $\beta$, and $\mathcal{H}(P)$ and $Z$ are the same as in Eq. (\ref{e-plink}). Once more, we use the Metropolis algorithm to estimate $R^N_A$.

Given the network reliability $R^N_A = p_{\rm BM}(A|A^O)$, the expected value of a property $X$
\begin{equation}
	\langle X \rangle = \sum_A X(A) \, R^N_A
	\label{e-reconstruct}
\end{equation}
over all possible networks $A$ is a better estimate of $X(A^T)$ than $X(A^O)$. We find that in many situations $R^N_{A^T} \gg R^N_{A^O}$ (Supporting Information, Fig. S13), which means that, presented only with an inaccurate observation $A^O$ (and with the knowledge about complex networks embodied in the stochastic block model family), our approach is remarkably able to identify that $A^T$ is a more likely network than the observation $A^O$ itself. This confirms that, even without knowing $A^T$, it is possible to estimate a property $X(A^T)$ better than just by measuring that property on $A^O$ (that is, better than assuming $X(A^T)=X(A^O)$). 

Since summing over all possible networks in Eq.~(\ref{e-reconstruct}) is prohibitive, we use the approximation $\langle X \rangle \approx X(A^R)$, where $A^R$ is the network that maximizes $R^N_A$ (in other words, $A^R$ is the maximum a posteriori estimate of $A$). The network $A^R$ is what we call a network reconstruction, and we claim that $X(A^R)$ is, in general, a better estimate of $X(A^T)$ than $X(A^O)$. In practice, we build reconstructions by heuristically maximizing $R^N_A$, starting from $A^O$ (Methods).

We test our network reconstruction approach by generating hypothetical observed networks $A^O$ from the true test networks $A^T$ described above. Each observation has a fraction of the true interactions removed (we call this fraction the observation error rate), and an identical number of random interactions added. In Fig.~\ref{f-reconstruct} we show the true air transportation network of Eastern Europe, as well as a hypothetical observation of this network (with an observation error rate of 20\%) and the corresponding reconstruction. The reconstruction has 13\% fewer missing and spurious interactions than the observation and, qualitatively, it appears that individual node properties (specifically, degree and betweenness centrality) are also better captured by the reconstruction.

However, from a systems perspective global network properties are more relevant than local node-level features. Therefore, the ultimate goal is to generate network reconstructions whose global properties are closer to those of the true network than those of the observations. To quantitatively investigate whether our approach accomplishes this aim, we calculate six network properties (static and dynamic) for observations and for the corresponding reconstructions of the air transportation network of Eastern Europe, and compute the relative error with respect to the true value. As we show in Fig.~\ref{f-netprop}, the reconstruction consistently improves the estimates of these properties. Only when the observed network contains less than 10\% of errors it is better, for a few of the properties, to use the observed network rather than the reconstruction. We obtain similar results for other networks and other network properties (Supporting Information, Figs. S8-S12).

\section{Application to a protein interaction network}
As we have discussed before, protein interaction networks are among the networks that may benefit the most from our approach. Ultimately, only experiments can prove our results useful; such experiments are, however, beyond the scope of this work. Nevertheless, here we show, however, how our approach can help in directing the effort to refine protein interaction data.

We consider the protein interaction network of yeast that Gavin {\it et al.} obtained using affinity purification and mass spectrometry (AP/MS) \cite{gavin06}. In AP/MS essays, a ``bait'' protein is used to detect ``prey'' proteins that interact with the bait directly or indirectly. Since bait and prey play different roles, we limit ourselves to the set of 991 proteins that are both viable baits and viable prey \cite{chiang09} (for example, we discard proteins that only appear as prey because prey-prey interactions cannot possibly be observed). We build a protein interaction network by connecting all pairs of proteins that are reported as a bait-prey pair at least once\footnote{Note that we do not advocate that this is the most appropriate procedure to analyze the structure of a protein interaction network (see \cite{chiang09} for a detailed discussion). Rather, we use this procedure because it enables us to test whether our algorithm can separate the least reliable and most reliable interactions.}. From this network, we obtain the reliability for all pairs of proteins.

We evaluate how successful our algorithm is by considering those proteins among the 991 in the network that have been used once, and only once, as bait (some proteins are used as bait in several independent essays). For a pair of these proteins $A$ and $B$, a link in the network can represent two distinct situations: (i) the interaction was observed once (with $A$ as bait and $B$ as prey but not the other way around, or vice versa); (ii) the interaction was observed twice (both with $A$ as bait and with $B$ as bait). Since these experimentally ``non-reproducible'' and ``reproducible'' interactions (3113 and 867 interactions in our network, respectively) are encoded identically in the network, it is interesting to see if our algorithm assigns lower reliability to the first and higher to the latter.

Remarkably, among the 100 interactions with the lowest link reliability according to our algorithm, only 5 are experimentally reproducible. Conversely, among the 100 interactions with the highest link reliability, as many as 65 are experimentally reproducible. The probabilities of observing by chance such a small number in the first case and such a large number in the latter case are $p_{\le} = 3 \times 10^{-6}$ and $p_{\ge} = 2 \times 10^{-20}$, respectively. Our approach is therefore successfully separating interactions that are likely to be spurious from those that are likely to be correct, without using any biophysical or biochemical information.

\section{Discussion}
We have shown that our network reconstruction method allows for a better characterization of network data sets, which will be particularly useful in data sets that we know contain many inaccuracies, such as protein interactomes. We have also shown that our approach reliably identifies missing and spurious interactions in complex networks, so that we can identify suspect interactions for further experimental probing.

Interestingly, our method can also guide new discoveries. If a given interaction between $i$ and $j$ truly exists but our approach predicts a very low reliability for the interaction (or vice versa), that means that the function of the interaction is very specific (since the interaction is rare among nodes that are otherwise similar to $i$ and $j$) and, therefore, functionally or evolutionarily important.

Finally, our approach is flexible enough to allow generalizations in several directions. Arguably the most important of these is the extension to arbitrarily sophisticated families of models. In particular, one could use models that are the ``product'' of a network model $M_n$ (probably a block model) and an error model $M_e$ that incorporates the relevant error structure (maybe another block model with non-uniform priors). The flexibility of our approach, along with its generality and its performance, will make it applicable to many areas where network data reliability is a source of concern.

\begin{materials}
\section{Outline of the reliability calculations}

Formally, a block model $M=(P, \mathbf{Q})$ is completely determined by the partition $P$ of nodes into groups and the matrix $\mathbf{Q}$ of probabilities of linkage between groups, so that Eq. (2) in the main text can be rewritten as
\begin{eqnarray}
	p_{\rm BM}(X=x|A^O) & = & \frac{1}{Z} \sum_{P \in \mathcal{P}} \int_{[0,1]^{G}} d\mathbf{Q} \; p(X=x|P, \mathbf{Q}) \times \nonumber\\
	& \times & p_{\rm BM}(A^O|P, \mathbf{Q}) \; p(P, \mathbf{Q}) \, ,
	\label{e-m-bayesBM}
\end{eqnarray}
where $\mathcal{P}$ is the space of all possible partitions of the network into groups, $G$ is the number of distinct group pairs, and $Z$ is a normalizing constant.

Within the family of stochastic block models, one can evaluate the likelihood of each model $M$ because the probability of any two nodes $i$ and $j$ being connected depends only on the groups to which they belong. We have that~\cite{holland83}
\begin{equation}
	p_{\rm BM}(A^O|P, \mathbf{Q}) = \prod_{\alpha \le \beta} Q_{\alpha \beta}^{l^O_{\alpha \beta}} \, (1 - Q_{\alpha \beta})^{r_{\alpha \beta} - l^O_{\alpha \beta}} \; ,
	\label{e-m-likelyhood}
\end{equation}
where $l^O_{\alpha \beta}$ is the number of links in $A^O$ between nodes in groups $\alpha$ and $\beta$ of $P$, and $r_{\alpha \beta}$ is the maximum number of such links (that is, the number of pairs of nodes such that one node is in $\alpha$ and the other is in $\beta$).

Note that, among all possible block models, there is at least one whose likelihood is 1, namely, the block model in which each node is in a different block and each $Q_{\alpha \beta}$ is 1 or 0 depending on whether the corresponding nodes are connected or not. This model contributes to $p(X=x|A^O)$ much more than most other models. However, there is only one such model (or very few), whereas there are many models with, for example, four blocks. This ``entropic'' term prevents overfitting of the network by very detailed (and ultimately uninformative) block models.

Using that $p(A_{ij}=1|P, \mathbf{Q})=Q_{\sigma_i \sigma_j}$ (where $\sigma_i$ is the module of node $i$ in partition $P$) and assuming no prior knowledge about the models (that is, $p(P, \mathbf{Q})={\rm const.}$), one can use Eqs.~(\ref{e-m-bayesBM}) and (\ref{e-m-likelyhood}) to obtain Eqs. (3)-(6) in the main text.

\section{Metropolis estimation of link and network reliability}

To estimate the link and network reliabilities given by Eqs.~(\ref{e-plink}) and (\ref{e-pnet}) we use the following procedure. We start by placing each of the $N$ nodes in a group, which we choose with uniform probability from a set of $N$ possible groups (that is, there are as many groups as nodes). In general, some of these $N$ groups will be empty after the initial node assignment.

At each step we select a random node and attempt to move it to a randomly selected group. This update scheme is appropriate because: (i) it results in an ergodic exploration of the space of possible partitions, and (ii) it satisfies detailed balance (since the probability of choosing a move and its reverse are identical). To decide whether we accept the move, we calculate the change $\Delta\mathcal{H}$ (Eq.~(\ref{e-H})): if $\Delta\mathcal{H}\le 0$, the change is automatically accepted; otherwise, the change is accepted with probability $\exp(-\Delta\mathcal{H})$.

The sampling procedure starts after an equilibration period, during which $\mathcal{H}$ decreases from an initial value to its equilibrium value. We sample the partition space by considering $S=10^4$ partitions, each one separated from the previous one by a number of steps that is large enough for the two partitions to be reasonably uncorrelated (as measured by the mutual information between partitions).

Because the link and network reliabilities are ensemble averages over independent partitions, it is straightforward to parallelize the algorithm so that the partitions are obtained concurrently. Therefore, given enough computational resources, the reliabilities can be calculated even for large networks (probably up to millions of nodes) in relatively short times.

\section{Benchmark algorithms for the identification of missing and spurious interactions}

The hierarchical random graph approach is described in detail in \cite{clauset08}. We use the implementation provided by the authors (available at http://www.santafe.edu/$\sim$aaronc/hierarchy/hrg\_20080819\_predictHRG\_v1.0.3.tgz), which we modified slightly to be able to study spurious as well as missing interactions.

We analyze three local algorithms: common neighbors, degree product, and Jaccard index (the last two in Supporting Information only). For each of these algorithms, the link ``reliability'' $R^L_{ij}$ is defined as follows (note that, for these approaches, the ``reliability'' is not a probability, but just a score that enables us to rank node pairs):

\begin{itemize}
 \item{Common neighbors:} $R^L_{ij}=\lVert \Gamma_i \cap \Gamma_j \rVert$, where $\Gamma_i$ is the set of neighbors of node $i$, and $\lVert \dots \rVert$ indicates the number of nodes in a set.

 \item{Degree product:} $R^L_{ij}=\lVert \Gamma_i \rVert \times \lVert \Gamma_j \rVert$.

 \item{Jaccard index:} $R^L_{ij}=\lVert \Gamma_i \cap \Gamma_j \rVert / \lVert \Gamma_i \cup \Gamma_j \rVert$.

\end{itemize}

\section{Heuristic network reconstruction}

The goal of the heuristic network reconstruction algorithm is to find $A^R = \operatorname{arg\,max}_A R^N_A$, where $R^N_A=p(A|A^O)$ is the reliability of network $A$ given observation $A^O$. Since exhaustive maximization of $R^N_A$ is not possible, we use the following heuristic method. Start by evaluating the link reliabilities $R^L_{ij}$ for all pairs of nodes in $A^O$; sort observed links ($A^O_{ij}=1$) by increasing reliability, and observed non-links ($A^O_{ij}=0$) by decreasing reliability. Then choose pairs of link/non-link in order: remove the link (which has a low reliability) and add the non-link (which has a high reliability), and accept the change if, and only if, $R^N_A$ increases. Repeat this procedure, going down the lists, until we reject five consecutive attempts to swap a link/non-link pair. At this point, reevaluate $R^L_{ij}$ and repeat the process. The algorithm stops when no link swaps are accepted.

\end{materials}

\begin{acknowledgments}
We thank L.A.N. Amaral, A. Arenas, A. D\'{\i}az-Guilera, J. Duch, K. Frank, R.D. Malmgren, P.D. McMullen, E.N. Sawardecker, D. Scholtens, and M.J. Stringer for useful comments and suggestions. R.G. and M.S.-P. gratefully acknowledge the support of NSF grant SBE-0830388. M.S.-P gratefully acknowledges the support of NIH grant CTSA-UL1RR025741.
\end{acknowledgments}


\end{article}

\clearpage

 \centerline{
 	\includegraphics*[]{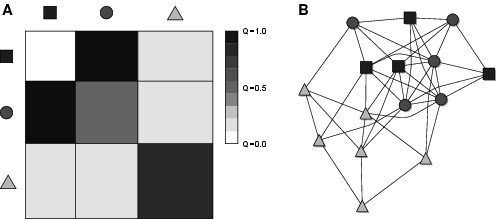}
 }
\centerline{Figure~\ref{f-blockmodel}}
\begin{figure}
	\centerline{
	}
	\caption{Stochastic block models. A stochastic block model is fully specified by a partition of nodes into groups and a matrix $\mathbf{Q}$ in which each element $Q_{\alpha\beta}$ represents the probability that a node in group $\alpha$ connects to a node in group $\beta$. {\bf A}, A simple matrix of probabilities $\mathbf{Q}$. Nodes are divided in three groups (which contain 4, 5, and 6 nodes, respectively) and are represented as squares, circles, and triangles depending on their group. The value of each element $Q_{\alpha\beta}$ is indicated by the shade of gray; for example, squares do not connect to other squares, and connect to triangles with small probability, but squares connect to circles with high probability. {\bf B}, A realization of the model in {\bf A}. In this realization, the number of links between the square and the triangle group is $l_{\Box \triangle}=4$, whereas the maximum possible number of links between these groups is $r_{\Box \triangle}=24$.}
	\label{f-blockmodel}
\end{figure}
\clearpage

\centerline{
	\includegraphics*[]{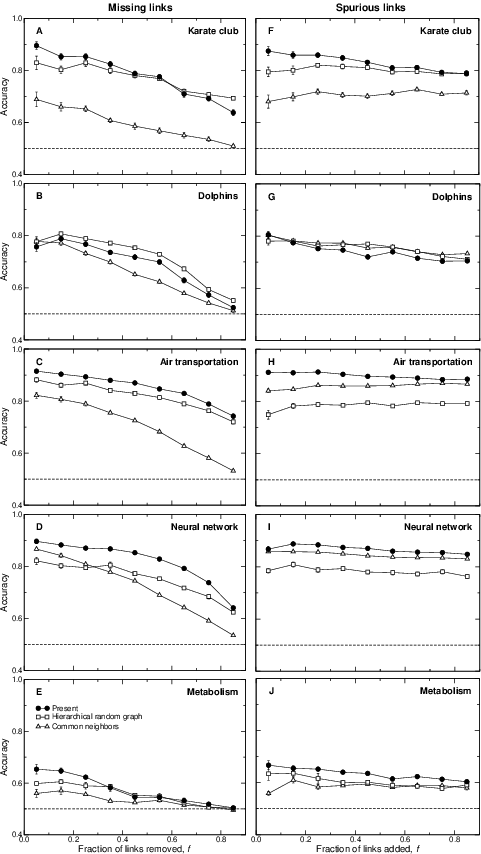}
}
\centerline{Figure~\ref{f-mis_bog_performance}}

\begin{figure}
	\caption{Identification of missing and spurious links. We compare the approach presented here (black circles), to the approach of Clauset {\it et al.}~\cite{clauset08} (white squares) and to a local algorithm based on the number of common neighbors between pairs of nodes~\cite{liben-nowell07,clauset08} (white triangles) (Methods; see Supporting Information for a comparison to other local algorithms).
	{\bf A-E} Missing links. For each true network $A^T$ we remove a fraction $f$ of its links to generate an observed network $A^O$, calculate the link reliability $R^L_{ij}$ for each pair of nodes, and rank pairs of nodes in order of decreasing reliability. Accuracy is calculated as the probability that a false negative (one of the links we removed, that is, $A^O_{ij}=0$ but $A^T_{ij}=1$) has a higher ranking than a true negative ($A^O_{ij}=0$ and $A^T_{ij}=0$). The dashed line indicates the baseline accuracy when false negatives and true negatives are randomly ranked.
	{\bf F-J} Bogus links. For each true network $A^T$ we add a fraction $f$ of links to generate an observed network $A^O$, calculate the link reliability $R^L_{ij}$ for each pair of nodes, and rank pairs of nodes in order of decreasing reliability. Accuracy is calculated as the probability that a false positive (one of the links we added, that is, $A^O_{ij}=1$ but $A^T_{ij}=0$) has a lower ranking than a true positive ($A^O_{ij}=1$ and $A^T_{ij}=1$). The dashed line indicates the baseline accuracy when false positives and true positives are randomly ranked.
	Scores for algorithms other than the present approach are obtained as described earlier \cite{clauset08} (Methods) and ties are randomly broken when necessary.
	}
	\label{f-mis_bog_performance}
\end{figure}
\clearpage

\centerline{
	\includegraphics*[]{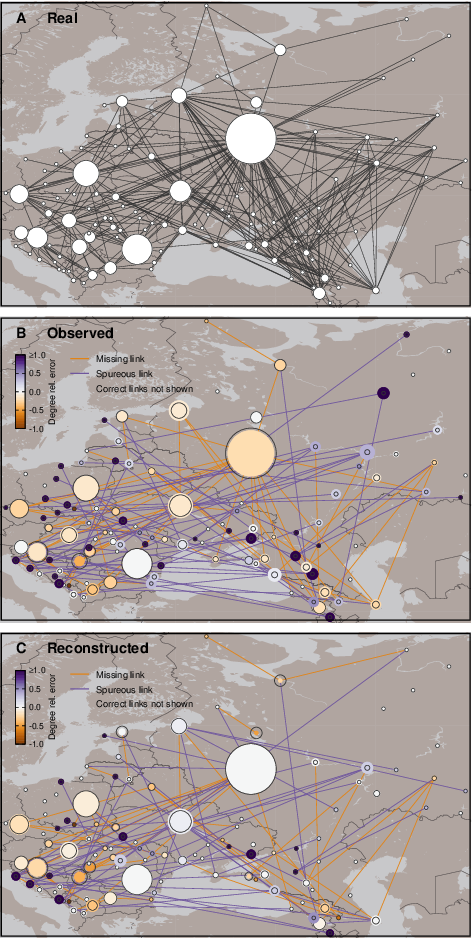}
}
\centerline{Figure~\ref{f-reconstruct}}
\begin{figure}
	\caption{Reconstruction of the air transportation network of Eastern Europe.
	{\bf A}, The true air transportation network. The area of each node is proportional to its betweenness centrality, with Moscow being the most central node in the network.
	{\bf B}, The observed air transportation network, which we build by randomly removing 20\% of the real links and replacing them by random links.
	{\bf C}, The reconstructed air transportation network that we obtain, from the observed network, applying the heuristic reconstruction method described in the text and methods.
	For clarity, in {\bf B} (respectively, {\bf C}) we do not depict the correct links, but only: (i) missing links in orange, which exist in the true network but not in the observation (reconstruction), and (ii) spurious links in blue, which do not exist in the true network but do exist in the observation (reconstruction).
	As in {\bf A}, the area of each node is proportional to its betweenness centrality, with the black circle representing the true betweenness centrality of each node. The color of each node represents the relative error in the degree of the node, with respect to the true degree.
	The observed network contains 60 missing and 60 spurious links, whereas the reconstruction only contains 52 of each (a 13\% improvement). In general, node degree and betweenness centrality are also better captured in the reconstruction.
	}
	\label{f-reconstruct}
\end{figure}
\clearpage

\centerline{
	\includegraphics*[]{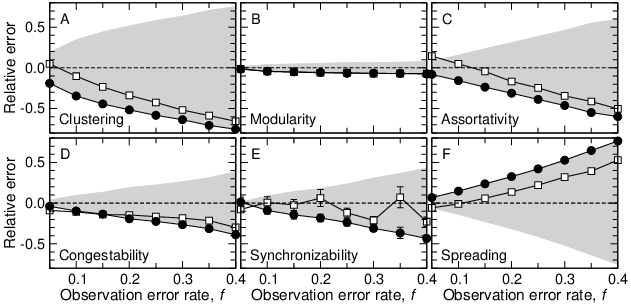}
}
\centerline{Figure~\ref{f-netprop}}
\begin{figure}
	\caption{Properties of observations and reconstructions of the air transportation network of Eastern Europe. In each case, the observation $A^O$ is generated from the true network $A^T$ by randomizing a fraction $f$ of its links. The reconstruction $A^R$ is generated from $A^O$ as described in the text and methods. For each property $X$, we calculate the relative error of the observation $(X(A^O) - X(A^T))/X(A^T)$ (black circles) and of the reconstruction $(X(A^R) - X(A^T))/X(A^T)$ (white squares). Symbols represent the mean over 25 repetitions, and the error bars indicate the standard error of the mean. The shaded region corresponds to the region with smaller relative error (in absolute value) than the observation, so that squares within the shaded region correspond to reconstructions that provide better network-property estimates than the observation itself. {\bf A-C}, Static properties: {\bf A}, Clustering coefficient~\cite{costa07}; {\bf B}, Modularity~\cite{costa07}; and {\bf C}, Assortativity~\cite{costa07}. {\bf D-F}, Dynamic properties: {\bf D}, Transportation congestability, that is, the maximum betweenness centrality in the network~\cite{guimera02a}; {\bf E}, Synchronizability, that is, the ratio between the largest eigenvalue and the smallest non-zero eigenvalue of the Laplacian matrix of the network~\cite{arenas08}; {\bf F}, Spreading threshold, that is, the ratio between the first and the second moments of the degree distribution~\cite{boguna02}.}
	\label{f-netprop}
\end{figure}

\end{document}